\documentclass[a4paper]{article}
\begin{document}
\title{Some questions concerning Bose-Einstein correlations in multiple
particle production processes.}
\author{K.Zalewski
\\ M.Smoluchowski Institute of Physics
\\ Jagellonian University, Cracow
\\ and\\ Institute of Nuclear Physics, Cracow}

\maketitle
\begin{abstract}
Most models of Bose-Einstein correlations in multiple particle production
processes can be ascribed to one of the following three broad classes: models
based on the original idea of the Goldhabers, Lee and Pais, hydrodynamic models
and string models. We present for discussion some basic questions concerning each
of these classes of models.
\end{abstract}

\section{Introduction}
Since the pioneering work of G. Goldhaber, S. Goldhaber, W. Lee and A. Pais,
published over forty years ago \cite{GGL} and known as the GGLP model,
Bose-Einstein correlations in multiple particle production processes have been
studied in hundreds of papers. Many references can be found in the recent review
articles \cite{WIH} and \cite{WEI}. These correlations have been popular in
particular for two reasons. They give impressive bumps in two-particle and many
particle distributions and, if the regime of Einstein's condensation can be
reached, there are even more spectacular phenomena waiting to be discovered
\cite{PR1}, \cite{PR2}, \cite{BZ1}, \cite{BZ2}. What is more, Bose-Einstein
correlations are believed to yield important information, which seems hard, if not
impossible, to obtain by other means. They have been used to find the sizes and
shapes of the regions where hadrons are produced, as well as to obtain detailed
information about the evolution in time of the hadron production processes (see
the reviews \cite{WIH}, \cite{WEI}).

There is little, doubt, however, that the problem is hard. As an example of a
sceptical opinion let me quote one of the creators of this field of research, G.
Goldhaber, who said at the Marburg conference in 1990: "{}What is clear is that we
have been working on this effect for thirty years. What is not as clear is that we
have come much closer to a precise understanding of the effect"{}. Everybody
agrees that the GGLP paper was very important and various extensions of the model
proposed there are still being used. There are, however, other approaches. The
GGLP model contains static sources of particles. In the more recent
"{}hydrodynamic"{} models the flow of the sources is of great importance. Another
very promising approach are string models, where the random phase assumption used
in previous models is not necessary and the description looks closer to QCD.

In the present paper we will characterize the GGLP models, the hydrodynamic models
and the string models, stressing in each case open and potentially important
problems.

\section{GGLP models}
Let us consider two identical bosons, e.g. two $\pi^+$ mesons, created: one  at
point $\vec{r}_1$ and the other at point $\vec{r}_2$. If the two bosons were
distinguishable, a crude approximation for the probability amplitude of observing
both of them at point $\vec{r}$ could be

\begin{equation}\label{}
  A_D = e^{i\phi_1 + i\vec{p}_1\cdot(\vec{r} - \vec{r}_1)}e^{i\phi_1 +
  i\vec{p}_2\cdot(\vec{r} - \vec{r}_2)}.
\end{equation}
Here the interaction between the two bosons is neglected, so that the two-particle
amplitude is a product of single particle amplitudes. Only the phase factors are
kept and each phase is the sum of the phase obtained by the boson at birth and of
the phase acquired while propagating with given momentum from the birth point to
point $\vec{r}$.

For identical bosons, however, this amplitude has not the right symmetry with
respect to the exchange of the two bosons and the least one must do is to
symmetrize it. Thus for identical bosons the corresponding approximation is

\begin{equation}\label{}
  A = \frac{1}{\sqrt{2}}e^{i(\phi_1 + \phi_2) +i(\vec{p}_1+\vec{p}_2)\cdot\vec{r}}
  \left(e^{-i(\vec{p}_1\cdot\vec{r}_1+ \vec{p}_2\cdot\vec{r}_2)} +
  e^{-i(\vec{p}_2\cdot\vec{r}_1+ \vec{p}_1\cdot\vec{r}_2)}\right).
\end{equation}
The probability distribution for momenta is proportional to

\begin{equation}\label{}
|A|^2 = 1 + \cos\left((\vec{p}_1 - \vec{p}_2)\cdot(\vec{r}_1 - \vec{r}_2)\right).
\end{equation}
In order to make use of this expression it is necessary to average it over the non
measured production points $\vec{r}_1,\vec{r}_2$. In the GGLP paper the averaging
was over the space distribution of sources $\rho(\vec{r}_1;R)\rho(\vec{r}_2;R)$,
where $R$ is a parameter with dimension length, which was interpreted as the
radius of the production region. Various generalizations, modifications and
extensions followed, but let us use this simple variant to make some general
remarks.

In the GGLP model the distribution for pairs of particle momenta depends only on
the momentum difference $\vec{q} = \vec{p}_1 - \vec{p}_2$. This is in violent
contradiction with the data, but GGLP found a clever way out. The distribution for
the unsymmetrized amplitude is flat. Therefore, the result can be just as well
interpreted as a prediction for the ratio of the actual momentum distribution to
the distribution for distinguishable bosons. Further we denote this ratio by
$R(\vec{p}_1,\vec{p}_2)$:

\begin{equation}\label{}
  R(\vec{p}_1,\vec{p}_2) = \frac{|A(\vec{p}_1,\vec{p}_2)|^2}
  {|A_D(\vec{p}_1,\vec{p}_2)|^2}
\end{equation}
Now, the momentum distribution is not assumed to be independent of the sum of the
momenta. It is enough to make the much weaker and more reasonable assumption that
the dependence on this sum can be factored out and cancels in the ratio. A well
known difficulty with this approach is that the distribution for distinguishable
$\pi^+$-s, say, cannot be obtained from experimental data without further
assumptions. GGLP assumed that the distribution for $\pi^+\pi^-$ pairs can be used
instead. There have been many other proposals (cf. e.g. \cite{HAY} and references
contained there), but none is fully satisfactory.

For any nonsingular averaging process the average cosine must be close to one for
$|\vec{q}| \approx 0$ and very small for large values of $|\vec{q}|$. Therefore,
the ratio $R(\vec{p}_1,\vec{p}_2)$ decreases, though not necessarily
monotonically, from values close to two for small values of $|\vec{q}|$ to values
close to one for large values of $|\vec{q}|$. This gives the characteristic bump
in $R(\vec{p}_1,\vec{p}_2)$ for small values of $\vec{q}^2$. If $R$ is the only
dimensional parameter available, the width of this bump must, for simple
dimensional reasons, be proportional to $R^{-2}$. Thus, the main qualitative
results of GGLP are much more general than their specific choices of the weight
functions $\rho(\vec{r};R)$. Nevertheless, they are not quite general.

It is well known from optics that, whether photons bunch or not, depends on the
type of source and not only on the fact that they are bosons. Photons can
antibunch just as well. In order to illustrate this point within the GGLP type
models, let us assume that the amplitude $A$ has an additional factor, which
equals one, if the product $(\vec{r}_1 - \vec{r}_2)(\vec{p}_1 - \vec{p}_2) > 0$
and minus one otherwise. This factor changes sign, when the momenta of the two
bosons are exchanged. Therefore, the squared modulus of the properly symmetrized
production amplitude is

\begin{equation}\label{}
|A|^2 = 1 - \cos\left((\vec{p}_1 - \vec{p}_2)\cdot(\vec{r}_1 - \vec{r}_2)\right).
\end{equation}
and we get a hole instead of the bump in the small $\vec{q}^2$ region. Admittedly
this model is not realistic. Its purpose is only to indicate a possibility. This
may be interesting in view of the LEP results concerning Bose-Einstein
correlations in $e^+e^-$ annihilations, where two $W$ bosons are simultaneously
produced. It seems that identical pions originating from the decay of a single $W$
exhibit the usual bump attributed to Bose-Einstein correlations, while these
correlations are absent, or very weak, for pairs of identical pions, when each
pion originates from a different $W$ \cite{ABB}, \cite{BAR}, \cite{ACC}. In the
GGLP model the bump results from the assumptions that pion pairs produced in
different pairs of points add incoherently. Mild modifications of this assumption
\cite{WIH}, \cite{WEI} can affect the size of the bump, but do not eliminate it.
It would be interesting to check, whether the GGLP assumptions could be modified
so as to predict the bump for some, but not for all pairs of identical mesons
produced in a multiple particle production event.

\section{Hydrodynamic models}
It is not possible to express the ratio $R(\vec{p},\vec{p}')$ in terms of the
single particle momentum distributions. In the hydrodynamic models, as well as in
GGLP models, one makes, however, an assumption, which makes it possible to express
this ratio in terms of the diagonal and off-diagonal terms of the single particle
density matrix in the momentum representation. It is convenient to formulate the
hydrodynamic models in terms of source functions $S(X,K)$. The source function
\cite{GKW}, \cite{PRA3} is related to the single particle density matrix in the
momentum representation by the formula

\begin{equation}\label{}
  \rho(\vec{p},\vec{p}') = \int e^{iqX}S(X,K)d^4X.
\end{equation}
In this formula

\begin{equation}\label{}
  K = \frac{1}{2}(p + p');\qquad q = p - p'.
\end{equation}
Here $K,q,p,p'$ are fourvectors, but in order to calculate the density matrix, we
need only their values corresponding to the momenta $p,p'$ being on their mass
shells. X is an integration variable, which is associated with the position of the
sources in space-time. The physical interpretation of X may be helpful, when
trying to find the source function. It is irrelevant for the calculation of the
density matrix, once the source function is known.

There is an infinity of different source functions, which all give the same
density matrix and consequently the same predictions for the ratio
$R(\vec{p},\vec{p}')$. For instance, one could put

\begin{equation}\label{}
  S(X,K) = W(\vec{X},\vec{K})\delta(X_0),
\end{equation}
where $W(\vec{X},\vec{K})$ is the well-know Wigner function satisfying the
relation

\begin{equation}\label{}
  \rho(\vec{p},\vec{p}') = \int e^{-i\vec{q}\cdot\vec{X}}W(\vec{X},\vec{K}) d^3X
\end{equation}
and

\begin{equation}\label{}
  \vec{X} = \frac{1}{2}(\vec{x} + \vec{x}').
\end{equation}
This source function gives the correct density matrix by construction, but it
corresponds to a most unlikely scenario, where all the particles are created
simultaneously at $X_0 = 0$. Since our aim is to find the correct density matrix,
this source function would be fine, in spite of the unlikely physical picture
attached to it. The problem is, however, that finding the Wigner function is not
any easier than finding the density matrix in the momentum representation. The
hope is that using a source function, which corresponds to a plausible scenario
for the production process, we will be able to use more efficiently what we know
about particle production in order to find the source function.

As an example of this approach let us consider the model reviewed in \cite{WIH}.
The source function is postulated in the form

\begin{eqnarray}\label{}
  S(X,K) = C m_T \cosh(y - \eta)\exp\left[\frac{m_T\cosh y \cosh \eta_t -
  r_T^{-1}XK_T\sinh \eta_t}{T}\right]*\nonumber\\ \exp\left[-\frac{r_T^2}{2R^2} -
  \frac{\eta^2}{2(\Delta\eta)^2} - \frac{(\tau -
  \tau_0)^2}{2(\Delta\tau)^2}\right].
\end{eqnarray}
In this formula

\begin{eqnarray}\label{}
  \eta = \frac{1}{2}\log\frac{t+z}{t-z};\qquad \eta_t = \eta_f\frac{r_T}{R};\qquad
  \tau = \sqrt{t^2 - z^2};\nonumber\\
  m_T^2 = m^2 + \vec{K}^2;\qquad r_T^2 = \vec{x}^2_T;
\end{eqnarray}
$z$ is parallel to $x_\|$ and $\vec{x}$ is parallel to $\vec{K}$. The source
function depends on six free parameters $(R, T, \eta_f, \Delta\eta, \tau_0,
\Delta\tau)$ and, moreover, contains the normalization constant $C$, but each
piece of the source function has a clear physical interpretation. Therefore,
fixing these parameters from the data yields directly interesting physical
information. We will illustrate this important point using a fit to the NA49 data
on Pb-Pb scattering at $158$ GeV/c per nucleon \cite{WIH}. The parameter $R$ is
the transverse radius of the tube, from which the final hadrons are emitted. The
result $R \approx 7$fm is about twice the radius resulting from from the known
radii of the lead nuclei. This is evidence for a significant transverse expansion,
before most of the hadrons are produced. The parameter $\eta_f$ governs the
transverse rapidity of the sources. The value obtained $\eta_f \approx 0.35$
corresponds to transverse velocities reaching the velocity of sound in the plasma
(1/3), which looks very reasonable. The parameter $T$ occurs as the temperature in
a Boltzmann type factor. Its fitted value $T \approx 130$fm is significantly lower
than the temperatures obtained, when fitting the chemical composition of the final
state hadrons, which indicates that during the expansion the stuff cools down.
Another interesting comparison is that of the fitted values $\tau_0 \approx 9$fm
and $\Delta\tau \approx 1.5$fm. The parameter $\tau_0$ is the typical time between
the moment of collision and the moment, when a hadron is produced. The parameter
$\Delta\tau$ is the duration of the time, when hadrons are produced. The fact that
$\tau_0 \gg \Delta\tau$ means that all the hadrons are produced in a short time
interval after a relatively long incubation time. Unfortunately, as stated by the
authors \cite{WIH}, the parameter $\Delta\tau$ is poorly constrained by the data,
so that this conclusion is not as solid as the others.

As seen from this example, given a model one can obtain from the data much
important information. An open problem is, however, how stable are these
conclusions when models change. As seen from the expression of the source function
in terms of the Wigner function, one can fit perfectly the data assuming that all
the particles are produced exactly simultaneously. It is just as easy to get a
perfect fit assuming that the particles are produced only on the surface of a
sphere, or only on the surface of a cube. These alternative models are so
implausible physically that there is little doubt they should be discarded. The
question is, however, how many physically plausible models can fit the data, while
giving completely different descriptions of the hadronization process?

\section{String picture}

The string model for Bose-Einstein correlations \cite{ANH},\cite{AND},\cite{ANR}
has not yet been developed to the point, where it could be compared quantitatively
with the data. It is, however, much more ambitious than the models described
above. Instead of phenomenological assumptions about sources and incoherence, it
gives a well defined amplitude for the production of particles with momenta
$\vec{p}_1,\ldots,\vec{p}_n$. This amplitude is a plausible approximation to QCD.
We will consider only the 1+1 dimensional version of the string model, which seems
to contain all the main ingredients of this approach, while it is much simpler
than the 3+1 dimensional version. In fact the 1+1 dimensional model has been
recently analytically diagonalized \cite{ANS}, though in the version without the
Bose-Einstein correlations.

Let us consider a final state consisting of hadrons with momenta $p_1,\ldots p_n$.
To this final state the model ascribes a polygon in the $(z,t)$ plane. The sides
of this polygon are the trajectories of the various partons existing between the
moment of $e^+e^-$ annihilation and the moments, when the hadrons are formed. The
partons are considered massless and moving with the velocity of light, therefore
the sides of the the polygon form angles $\pm 45^\circ$ with the $t$ and $z$ axes.
Let us put the $e^+e^-$ annihilation point at the origin of the coordinate system.
At this point two partons, a quark and an antiquark, are formed flying along the
$z$ axis, away from each other. Their trajectories form the first two sides of the
polygon, both starting at the origin, one going to the right and upwards, the
other going to the left and upwards. The partons are end points of a
colour-string. Thus the string sweeps the surface of the polygon. The energy of
the string $E$ is connected to its length $L$ by the formula $E = \kappa L$, where
$\kappa$ is a constant known as the string tension. Thus, while the quark and the
antiquark fly away from each other and the string expands, there is a force
reducing  the energy of the two partons and finally the directions of their
motions get reversed starting another pair of the sides of the polygon. In the
meantime the string at any point between the endpoints can break producing a quark
and an antiquark, which form another pair of sides of the polygon. Since any
segment of the string has a quark at one end and an antiquark at the other, it is
easily checked that, except for the original two partons,  all the quarks fly in
one direction and all the antiquarks in the other. From time to time a quark meets
and antiquark. Then the two form a hadron with two-momentum equal to the sum of
the two-momenta of the two meeting partons. Thus the polygon contains the
following elements: the vertex at the origin, two turning points of the original
partons, $n$ vertices, where the $n$ hadrons were formed and $n-1$ vertices, where
the string broke. One finds that the two-momentum of a hadron is determined by and
determines the lengths of the two sides of the polygon adjacent to the vertex
where the hadron was formed.

The probability amplitude for producing the state $(p_1,\ldots,p_n)$ depends on
the area $A$ of the polygon

\begin{equation}\label{amplstr}
M(p_1,\ldots,p_n) \sim e^{i\xi A}.
\end{equation}
The imaginary part of $\xi$, $ib$, gives the probability distribution well-known
from the LUND model

\begin{equation}\label{}
|M(p_1,\ldots,p_n)|^2 \sim e^{-bA}.
\end{equation}
The imaginary part, believed to be close to the string tension $\kappa$, is
important for the description of the Bose-Einstein correlations. In order to
describe these Bose-Einstein correlations the amplitude (\ref{amplstr}) is
symmetrized very much like in the GGLP approach and a qualitatively satisfactory
description of the correlations is obtained. Since, however, the amplitude being
symmetrized is not the GGLP one, there are some significant new points.

Symmetrization means summing over all the permutations of identical hadrons. Let
us concentrate on an exchange of two $\pi^+$ mesons. Each of them is produced at
some hadronization vertex of the polygon. In order to perform the exchange, one
has to cut off the two pionic vertices together with their adjacent sides of the
polygon and to glue them back, pion one in the position of pion two and pion two
in the position of pion one, so as to obtain again a closed polygon. The new
polygon has in general a different area $A' \neq A$. If the new area is much
larger than the area before the exchange, the contribution from the interference
with the permuted amplitude is negligible. One reason, familiar from the GGLP
model, is that the relative phase is a rapidly varying function of momentum. The
new fact is, however, that the modulus of the permuted amplitude is additionally
suppressed by the factor $e^{-\frac{b}{2}(A-A')}$. In order to obtain small
changes of the area, it is advantageous to exchange pions, which are close to each
other counting along the perimeter of the polygon, or equivalently, which have
similar momenta. This is the reason for the familiar bump for $p_1 \approx p_2$

Let us quote two interesting qualitative predictions of this model. There should
be a difference between the Bose-Einstein correlations for pairs of charged pions,
say $\pi^+\pi^+$, and corresponding correlations for pairs of neutral pions. The
reason is that two $\pi^0$-s can be formed at two adjacent hadronisation vertices
of the polygon, while two $\pi^+$-s cannot. There should also be a difference
between the Bose-Einstein correlations for pairs of mesons originating from the
same string and pairs of mesons originating from different strings. For the latter
situation the present model is clearly not applicable. It has been suggested
(\cite{AND2} and references quoted there) that perhaps there are no correlations
between mesons from different strings, which would explain the observations of
Bose-Einstein correlations for pions from decays of pairs of $W$ bosons.

An interesting question raised by Bowler is the relation of the string model to
the GGLP model. Bowler proposed \cite{BOW} a model closely related to the GGLP
model, which looks very similar to the string model and, according to Bowler,
gives also very similar predictions. The model used by Bowler contains a
distribution of sources, which depends not only on space-time points, as in the
GGLP paper, but also on the momenta of the produced particles. Such models have
become popular after the work Yano and Koonin \cite{YAK} and are sometimes called
Yano Koonin models.  It is not clear, however, what constraints must be imposed on
the distributions, in order to make them consistent with quantum mechanics.
Therefore, occasionally this approach gives inconsistent results \cite{MKFW}.
Bowler's model deviates in two ways from the string model. It yields the
probability amplitude as an exponential in the area, but this is a modified area,
where certain regions are counted more than once. Moreover, in both models, in
order to get the inclusive $k$-particle distribution it is necessary to integrate
out the momenta of the remaining particles, but in Bowler's model the integration
region is different from that used in the string model. According to some
unpublished numerical calculations by Bowler the two deviations nearly cancel.


\begin{thebibliography} {99}
\bibitem{GGL}G. Goldhaber, S. Goldhaber, W. Lee and A. Pais, {\it Phys. Rev.}
{\bf 120}(1960)300.
\bibitem{WIH}U.A. Wiedemann and U. Heinz,{\it Phys. Rep.} {\bf 319}(1999)145.
\bibitem{WEI}R.M. Weiner, {\it Phys. Rep.} \textbf{327}(2000)250.
\bibitem{PR1}S. Pratt, {\it Phys. Letters} {\bf B301}(1993)159.
\bibitem{PR2}S. Pratt, {\it Phys. Rev.} {\bf C50}(1994)469.
\bibitem{BZ1}A. Bialas and K. Zalewski, {\it Eur. Phys. J.} {\bf C6}(1999)349.
\bibitem{BZ2}A. Bialas and K. Zalewski, {\it Phys. Rev.} {\bf D59}(1999)097502.
\bibitem{HAY}S. Haywood, {\it Where are we going with Bose-Einstein -- a mini
review} RAL report January 6-th 1995.
\bibitem{ABB}G. Abbiendi et al., {\it Eur. Phys. J.} {\bf C8}(1999)559.
\bibitem{BAR}R. Barate et al., {\it Phys. Letters} {\bf B478}(2000)50.
\bibitem{ACC}M. Acciarri et al., {\it Phys. Letters} {\bf B493}(2000)233. 
\bibitem{GKW}M. Gyulassy, S.K. Kauffmann and L.W. Wilson, {\it Phys. Rev.} {\bf
C20}(1979)2267.
\bibitem{PRA3}S. Pratt, {\it Phys. Rev. Letters} {\bf 53}(1984)1219.
\bibitem{ANH}B. Andersson and W. Hofmann, {\it Phys. Letters} {\bf 169B}(1986)364.
\bibitem{AND}B. Andersson, {\it Acta Phys. Pol.} {\bf B29}(1998)1885.
\bibitem{ANR}B. Andersson and M. Ringn$\grave{e}$r, {\it Nucl. Phys} {\bf
B513}(1998)627.
\bibitem{ANS}B. Andersson and F. S\"odergerg, {\it Eur. Phys. J.} {\bf C16}{2000}.
\bibitem{AND2}B. Andersson, Moriond 2000.
\bibitem{BOW}M.G. Bowler, {\it Phys. Letters} {\bf B185}(1987)205.
\bibitem{YAK}F. Yano and S. Koonin, {\it Phys. Letters} {\bf B78}(1978)556.
\bibitem{MKFW}M. Martin, H. Kalechofsky, P. Foka and U.A. Wiedemann, {\it Eur.
Phys. J} {\bf C2}(1998)359.
\end{thebibliography}
\end{document}